\documentclass[aps,prb,twocolumn,showpacs,groupedaddress]{revtex4}
\usepackage{amsbsy,amssymb,amsmath,bm,amstext}
\usepackage{epsfig,wrapfig}
\usepackage{epstopdf}
\usepackage{array}
\usepackage[toc]{appendix}
\newcolumntype{L}[1]{>{\raggedright\let\newline\\\arraybackslash\hspace{0pt}}m{#1}}
\newcolumntype{C}[1]{>{\centering\let\newline\\\arraybackslash\hspace{0pt}}m{#1}}
\newcolumntype{R}[1]{>{\raggedleft\let\newline\\\arraybackslash\hspace{0pt}}m{#1}}

\input{epsf}
\begin{document}

\title{Electrodynamics of a planar Archimedean spiral resonator}
\author{N. Maleeva$^1$, N. N. Abramov$^1$, A. S. Averkin$^1$, M. V. Fistul$^{1,2}$, A. Karpov$^2$, A. P. Zhuravel$^3$, and A. V. Ustinov$^{2,4}$}
\affiliation{$^1$National University of Science and Technology "MISIS", Laboratory for Superconducting Metamaterials, Leninsky Prospect, 119049, Moscow, Russia\\
$^2$ Theoretische Physik III, Ruhr-Universit\"at Bochum, D-44801 Bochum, Germany\\
$^3$B. Verkin Institute for Low Temperature Physics and Engineering, National Academy of Sciences of Ukraine, Kharkov, 61103, Ukraine\\
$^4$Physikalisches Institut, Karlsruhe Institute of Technology (KIT), Karlsruhe, 76131, Germany}
\date{\today}

\begin{abstract}

We present a theoretical and experimental study of electrodynamics of a planar spiral superconducting resonator of a finite length. The resonator is made in the  form of  a monofilar Archimedean spiral.  By making use of a general model of inhomogeneous alternating current flowing along the resonator and specific boundary conditions on the surface of the strip, we obtain analytically the frequencies $f_n$ of resonances which can be excited in such  system. We also calculate corresponding inhomogeneous RF current distributions $\psi_n (r)$, where $r$ is the coordinate across a spiral.  We show that  the resonant frequencies and current distributions are well described by simple relationships $f_n=f_1 n$, and $\psi_n(r)\simeq \sin[\pi n (r/R_e)^2]$, where $n=1,2...$, and $R_e$ is the external radius of the spiral.
Our analysis of electrodynamic properties of spiral resonators' is in good accord with direct numerical simulations and measurements made using specifically designed magnetic probe and laser scanning microscope.
\end{abstract}


\maketitle

\section{Introduction}

Magic experiments by Nikola Tesla, that generated so much of excitement in the public of the $Belle Epoque$, had a solid base in research and patent activity. A spiral resonator appeared first in one of Tesla patents as early as 1897, as a tool for wireless RF power transfer.\cite{Tesla1,Tesla} Despite a wide use of the planar Archimedean spiral as antenna along the XX century,\cite{antena1, antena2} a possibility to develop an analytical model of electrodynamics of a planar Archimedean spiral resonator and its inner resonance modes appeared to be elusive and was not fully explored. Our own attention to this topic is related to the development of metamaterials.\cite{SMM}

Metamaterial is an artificially tailored media showing unusual electrodynamic properties. It is based on the use of compact magnetic and electric resonant elements, so-called meta-atoms.\cite{MetaMatGen} The electrical and magnetic meta-atoms are sub-wavelength size micro-resonators that couple primarily to either electric or magnetic field of the incoming electromagnetic wave. At certain frequencies, close to resonances of both meta-atoms' types, the effective permeability and permittivity of such a media may become negative, resulting in effective negative index of refraction of the media. This effect is giving a multitude of opportunities for inventing non-trivial new optics, as pointed out in early theoretical work by V.G. Veselago from 1967\cite{Veselago} and more recent reviews.\cite{PL, Zheludev, Philipp}

The first proposed magnetic meta-atoms were split-ring resonators.\cite{SMM,Pendry,Shelby,MMSRR} The resonance frequency of such resonators is determined by the ratio between the width of the gap  $l$ and the size $R$ of the ring, somewhat limiting the minimum  ratio $R/\lambda$, where $\lambda$ is the wavelength. The usage of planar spiral resonators was suggested in order to radically reduce the resonator size relative to the wavelength.\cite{SR1, Anlage2, SpiralArray} Particularly, a spiral resonator behaves as a distributed resonator with multiple resonance modes, and couples primarily to the magnetic component of RF field perpendicular to spiral plane, in a way suitable for magnetic meta-atoms.

In this work, we develop an analytical model of Archimedean spiral resonator inner modes and verify it by experiments and detailed numerical simulations.

In a very simplified approach, the spiral resonator may be considered as straight-line resonator of a length L, rolled in a spiral (Fig. \ref{scheme}). This view leads to an assumption that the spiral resonance frequencies are determined as the resonances in one-dimensional transmission line resonator with the length $L$ with identical "open circuit" boundary conditions at both ends of transmission line 
\begin{equation}\label{qualit_freq}
f_n=n f_1=n \frac{c}{2L},
\end{equation}
where $n=1,2...$ is the resonance number, $c$ is the speed of light, and in our case $L=2\pi R_e^2/d$, where $R_e$ is the external radius of the spiral and $d$ is  the spiral period. The amplitude of a standing wave as a function of the distance $l$ along the transmission line rolled in spiral, assuming the line to be uniform, is expected to be $I_n = I_0 \sin(n \pi \frac{l}{L})$. When the coordinate is expressed in terms of the radius $r$, the waveform at a resonance frequency is
\begin{equation}\label{qualit}
I_n(r)=I_0 \sin\Big(n \pi \Big(\frac{r}{R_e}\Big)^2\Big).
\end{equation}
While being simple and feasible, this approach nevertheless neglects the spiral geometry and interaction between the spiral turns.


Indeed, the above simple model is not always correct. For example, in contrast to Eq. (\ref{qualit_freq}), for a ring-shaped spiral resonator the resonance frequencies expression is $f_n\approx(2n-1)f_1$.\cite{Ustinov, ourarticle} Similarly, if the line is rolled in a ring-shaped spiral resonator\cite{ourarticle} the RF current amplitude distribution at resonance frequency is given by a more intricate expression then Eq.(\ref{qualit}).
From this point of view, it appears relevant to study in more detail the resonance frequencies and the corresponding current distributions of a resonator made in the form of a planar Archimedean spiral.

In the following, we provide an analytical solution of Archimedean spiral resonator problem, obtaining both the resonance frequencies $f_n$ and the corresponding current distributions in a resonator. We compare our analytical solution with the direct numerical simulation of electrodynamic properties of a spiral resonator excited by externally applied RF magnetic field. We present also an experimental study of microwave resonances of a single Archimedean spiral resonator by using both RF transmission measurements and two different spatially-resolved probing techniques. From our experiments, we are able to extract the resonance frequencies and distributions of current and of the magnetic field near the spiral at these frequencies. Our experimental results firmly support the proposed electrodynamic model and analysis.

The paper is organized as follows.
In section II, we propose an analytical model of the planar Archimedean spiral resonator of a finite length and obtain the integro-differential equation determining the resonance frequencies $f_n$ and the corresponding current distributions. We provide the approximate analytic solution of this equation, and calculate the resonance frequencies and current distributions. In section III, we present an experimental study of planar Archimedean spiral resonators. The  two different resonators were studied. The resonance frequencies are determined through the measurement of the frequency dependent transmission  of an externally applied RF signal, i.e. $S_{21}(f)$. The experimental current and the magnetic field distributions at the resonances were obtained by two methods: Laser Scanning Microscopy (LSM) for a resonator fabricated from superconducting Nb film and Magnetic Probe Scanning (MPS) for a resonator made of Cu.  In section IV,  we present the direct numerical simulations of electrodynamic properties, in particular, the resonance frequencies and the current distributions in a spiral resonator exited by an externally applied microwave radiation. Section V contains conclusions.

\section{Electrodynamics of a planar spiral resonator}

\begin{figure}[tbp]
\includegraphics[width=3.2in]{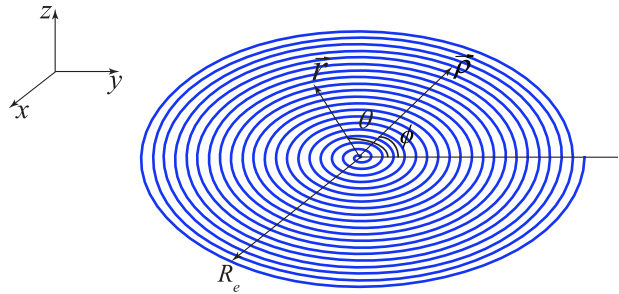}
\caption{A sketch of monofilar planar Archimedean spiral resonator. Polar coordinates $\{\rho, \varphi\}$ and $\{r, \theta\}$ determine the coordinates of the point on the spiral and the observation point in the plane of the spiral, accordingly. Here $R_e$ is an external radius of the spiral.}
\label{scheme}
\end{figure}

We consider the electrodynamics of a monofilar Archimedean spiral resonator of a finite length with $N$ densely packed turns. In this case the number of turns $N\gg 1$. The shape of the Archimedean spiral is described by equation written  in polar coordinates as
\begin{equation}
\rho(\varphi)=R_e(1-\alpha\varphi),\label{SE}
\end{equation}
where $\varphi$ is the polar angle, $\rho$ is the polar radius varying from $0$ to the external radius of the spiral $R_e$, and the parameter $\alpha =\frac{d}{(2\pi R_e)}=\frac{1}{(2\pi N)} \ll 1$, and $d$ is the distance between adjacent turns. The schematic view of such a spiral resonator is shown in Fig. \ref{scheme}.

In order to obtain electrodynamic properties of a spiral resonator and, in particular, the resonance frequencies $f_n$, we use the method proposed in Refs. \cite{Kogan,Silin} for a resonator made in the form of a helical coil. A similar approach has been used also in Ref. \cite{ourarticle} in order to describe electrodynamics of a ring-shaped spiral resonator. We assume that an alternating inhomogeneous current flows \emph{along} the spiral. We also neglect current inhomogeneity inside of the spiral line. The coordinate and the time dependent radial and angular components of a vector-potential in cylindrical coordinates can be presented in the following form: \cite{ourarticle}
\begin{equation}
\begin{gathered}
A_r(t, z, r, \theta)=\frac{\mu_0 I e^{i\omega t}}{(4\pi)^2} \int_{0}^{R_e} d \rho \psi(\rho)\cdot
\\
\cdot \int_0^\infty dx\frac{x e^{-\sqrt{x^2-k^2}|z|}}{\sqrt{x^2-k^2}} J_1[x\rho ] J_1[xr] \label{RproectionA-2}
\end{gathered}
\end{equation}
and
\begin{equation}
\begin{gathered}
A_\theta(t, z, r, \theta)=\frac{\mu_0 I e^{i\omega t}}{(4\pi)^2} \int_{0}^{R_e} d \rho \frac{\rho \psi(\rho)}{R_e\alpha}\cdot
\\
\cdot \int_0^\infty dx\frac{x e^{-\sqrt{x^2-k^2}|z|}}{\sqrt{x^2-k^2}} J_1[x\rho ] J_1[xr], \label{ThetaproectionA-2}
\end{gathered}
\end{equation}
where $I$ is the maximum value of the current excited in the spiral, $k=\omega/c$ is the wave vector, $z$ is a coordinate perpendicular to the spiral plane, $\psi(\rho)$ describes normalized inhomogeneous current distribution \emph{across} the spiral in the radial direction, and $J_1$ is the Bessel function of the first kind.\cite{AS}

In order to obtain the resonance frequencies $f_n$ and the corresponding current distributions $\psi_n(\rho)$, we use a specific boundary condition, i.e. we require the component of an electric field parallel to the strip surface to be equal to zero. This condition is written as:
\begin{equation}
R_e \alpha E_r+r E_\theta|_{z=0}=0.\label{Condition}
\end{equation}
The radial component $E_r$ and the angular component $E_\theta$ of the  electric field are expressed through corresponding components of the vector-potential as:

\begin{equation}
E_r=\frac{1}{i\omega \epsilon_0 \mu_0}\frac{d}{dr}\left[\frac{1}{r}\frac{d}{dr}(rA_r)\right] \label{Relectricfield}
\end{equation}
and
\begin{equation}
E_\theta=-i\omega A_\theta. \label{Thetaelectricfield}
\end{equation}
The Eqs. (\ref{RproectionA-2}) and (\ref{ThetaproectionA-2}) can be simplified by using the following approximation: the wave vector $k$ is much smaller than a typical inverse size of inhomogeneities in current distribution $\psi(\rho)$, i.e. $k<<1/R_e$.

With this approximation both components of the vector-potential in the plane of the spiral ($z=0$)  are written as:
\begin{equation}
\begin{gathered}
\left.A_r\right|_{z=0}=\frac{\mu_0 I e^{i\omega t}}{(4\pi)^2}\int_{0}^{\infty}dz \int_{0}^{R_e} d \rho \psi(\rho) \frac{1}{r^2}\cdot
\\
\cdot \int_{0}^{\infty}dx ~e^{-\frac{zx}{r}} x J_1[\frac{\rho}{r}x]J_1[x] \label{RproectionA-3}
\end{gathered}
\end{equation}
and
\begin{equation}
\begin{gathered}
\left.A_\theta\right|_{z=0}=\frac{\mu_0 I e^{i\omega t}}{(4\pi)^2 R_e \alpha}
\int_{0}^{\infty}dz \int_{0}^{R_e} d \rho \psi(\rho)\frac{\rho}{r^2} \cdot
\\
\cdot \int_{0}^{\infty}dx ~x e^{-\frac{zx}{r}}  J_1[\frac{\rho}{r}x]J_1[x]. \label{ThetaproectionA-3}
\end{gathered}
\end{equation}

Introducing the new variables $\rho=R_e e^{-\tau}$ and $r=R_e e^{-\xi}$ the boundary condition (\ref{Condition}) is written in the form of an integro-differential equation [see the details in Appendix]
\begin{equation}\label{Int-diff-Equation}
\begin{gathered}
\int_{0}^{\infty}dz \int_{0}^{\infty} d \tau \psi(\tau) \Big(e^{3\xi}K''_\xi(\xi-\tau)+
\\
+2e^{3\xi}K'_\xi(\xi-\tau) + \frac {\omega^2{R_e}^2}{c^2\alpha^2}e^{\xi-\tau}e^{-\xi}K(\xi-\tau) \Big)=0,
\end{gathered}
\end{equation}
where the kernel $
K(\xi-\tau)=e^{-(\tau-\xi)}\int_{0}^{\infty}dx x e^{-\frac{zx}{r}} J_1[e^{-(\tau-\xi)}x]J_1[x]
$. Since the kernel $K(\xi-\tau)$ resembles the $\delta$-function we apply a local approximation and obtain the differential equation for the current distribution $\psi(\xi)$:
\begin{equation} \label{integral-differential eq}
\begin{gathered}
\int_{0}^{\infty}dz \bigg[ \psi'(0)e^{3\xi}K (\xi)+
\\
+\int_{-\xi}^{\infty}du K(u) \Big(e^{3\xi}\psi''_\xi(\xi)+2e^{3\xi}\psi'_\xi(\xi) +
\\
+\frac {\omega^2{R_e}^2}{c^2\alpha^2} e^{-u}e^{-\xi}\psi(\xi)\Big)\bigg]=0.
\end{gathered}
\end{equation}
Here, we have taken into account that $\left.\psi(\tau)\right|_{\tau=0,\infty}=0$ and $K(\infty)=0$.


Introducing a new variable  $v=(r/R_e)^2=e^{-2\xi}$ and explicitly calculating the integral over $z$ we obtain:
\begin{equation} \label{integral-differential eq_modified}
\begin{gathered}
-2\frac{K (v)}{g_1(v)} \psi'_v (1)v^{-2} + 4 \psi''_v(v)+
\\
+\frac {\omega^2{R_e}^2}{c^2\alpha^2} \frac{g_2(v)}{g_1(v)}\psi(v)=0.
\end{gathered}
\end{equation}
Here, the expressions for the kernel $K(v)$, and functions $g_1(v)$ and $g_2(v)$ are written as $K(v)=R_e v\int_0^\infty dx J_1[{v}^\frac{1}{2} x]J_1[x]$, $g_1(v)=R_e v^{\frac{1}{2}} \int_{\sqrt v}^{\infty} du  \int_0^\infty dx J_1[ux]J_1[x]$ and
$g_2(v)=R_e v^{\frac{1}{2}} \int_{\sqrt v}^{\infty} du u^{-1} \int_0^\infty dx J_1[ux]J_1[x]$.

The solution of Eq. (\ref{integral-differential eq_modified}) can be obtained as following: first, we notice that with  a good accuracy the ratio $\frac{g_2(v)}{g_1(v)}$ is equal to 1. Moreover, as a first approximation we neglect a ratio $\frac{K(v)}{g_1(v)}$. In this case Eq.(\ref{integral-differential eq_modified}) is simplified  to such form:
\begin{equation}\label{integral-differential eq_simple}
\psi''_v (v) +\frac {\omega^2{R_e}^2}{4c^2\alpha^2}\psi(v)=0.
\end{equation}
A solution of Eq. (\ref{integral-differential eq_simple}) is $\psi(v)=\sin\Big(\frac {\omega{R_e}}{2c\alpha}v\Big)$, where $v=r^2$, and therefore, we arrive on Eq. (\ref{qualit}) obtained by qualitative analysis.

In order to improve such approximation we  look  for a solution of Eq.(\ref{integral-differential eq_modified}) in the following form:
\begin{equation}\label{sin_solution}
\psi(v)=\sum_n A_n\sin\pi nv.
\end{equation}
Substituting (\ref{sin_solution}) into (\ref{integral-differential eq_modified}) we obtain:
\begin{equation}
\begin{gathered}
-2\pi\beta_m\sum_n A_n n (-1)^n-2\pi^2A_m m^2+
\\
+\frac {\omega^2{R_e}^2}{2c^2\alpha^2} A_m=0,
\end{gathered}
\end{equation}
where $\beta_m=\int_0^1dv \frac{K (v)}{g_1( v)}v^{-2}\sin\pi m v$. The coefficients $A_m$ are expressed in the following form:
\begin{equation}\label{A-solution}
A_m=\frac{2\pi\beta_m S}{-2\pi^2m^2+\frac {\omega^2{R_e}^2}{2c^2\alpha^2}}.
\end{equation}
Here, $S=\sum_n A_n n (-1)^n$. The coefficients $A_n$ were calculated for different resonant modes. The distributions of $A_n$ for $n=1...10$ are shown in Fig. 2. One can see  that the main contribution in the form $\psi(v)=\sum_n A_n\sin\pi nv$ makes a summand coinciding with the Eq. (\ref{qualit}).
\begin{figure}[tbp]
\includegraphics[width=3.2in]{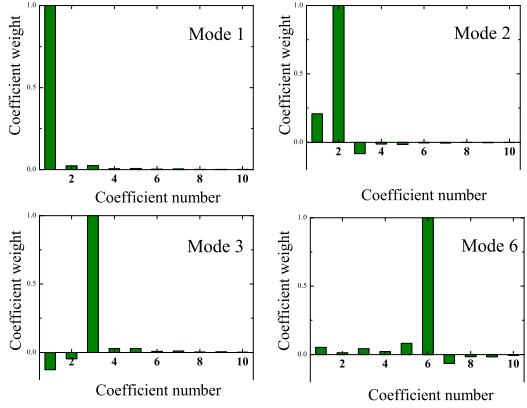}
\caption{Analytically predicted RF current amplitude distributions across the spiral  at the first, second, third, and sixth resonance frequencies. The distributions are presented as the coefficients $A_n$ of Eq. (\ref{sin_solution}). Each diagram shows the contribution of components with the corresponding resonance numbers into the waveform given by the "intuitive" solution (Eq.(\ref{qualit})). One can conclude that the most significant part of the standing wave profile can be explained by the solution of the simplified Eq. (\ref{integral-differential eq_simple}), amounting for about $90-80 \%$ of RF current at each mode.}
\label{Coeff}
\end{figure}
\begin{figure}[tbp]
\includegraphics[width=3.2in]{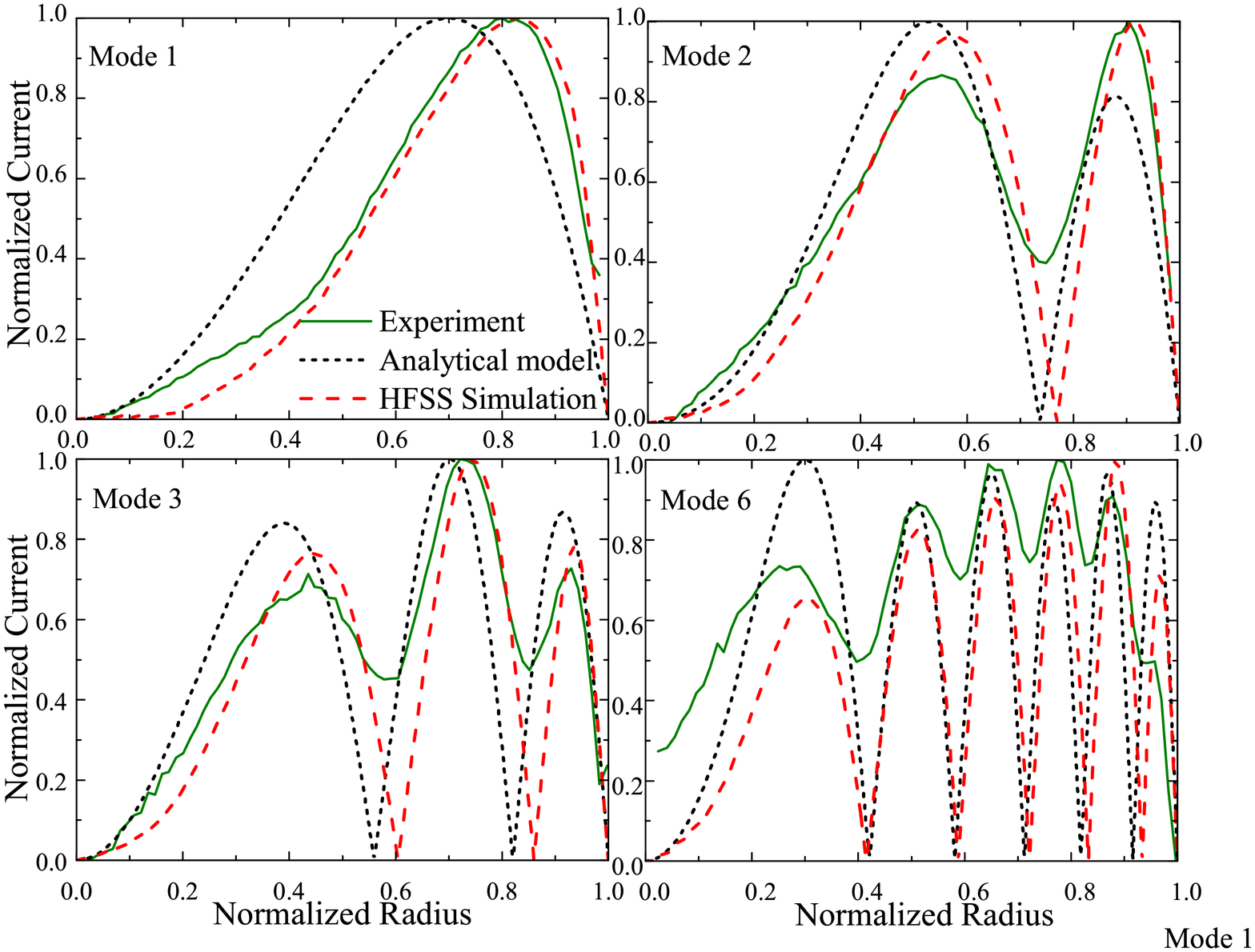}
\caption{Experimental and theoretical RF current distributions along the spiral radius for the first, second, third and sixth inner resonance mode of a superconductive Nb spiral (SR2). The data is measured with Scanning Laser Microscope (LSM) (solid line). Although smoothed by the thermal dissipation in the substrate, experimental data is visibly in a good agreement with analytical model (dotted line) and HFSS simulation (dashed line).  The spiral dimensions are $R_e=1.5$ mm, $N=75$, $d=0.02$ mm , and the substrate is 0.35 mm thick Si.}
\label{Currents}
\end{figure}
The resonance frequencies of the spiral are determined by a transcendent equation:

\begin{equation}\label{Frequency}
\pi=\sum_m\frac{(-1)^m m\beta_m}{-m^2+\frac {\omega^2{R_e}^2}{4\pi^2c^2\alpha^2}}.
\end{equation}

Next, we numerically solve the Eq. (\ref{Frequency}) and, therefore, obtain the resonance frequencies $f_n$ for two spiral resonators: the first one (SR1) had dimensions $R_e=16.25$ mm, $N=23$, $d=0.7$ mm, and the second one (SR2) had dimensions $R_e=1.5$ mm, $N=75$, $d=0.02$ mm. In the Table \ref{freq_table} (columns 3 and 6) the resonance frequencies $f_n$ for ten modes ($n=1...10$) are presented. Note here that  in experiment (see Section III) and in simulation (see Section IV) the spirals were placed at the substrate with dielectric constant $\epsilon_r$. It is known that the effective dielectric constant for a structure at the interface of dielectric and air is approximately $\epsilon_{eff}=\frac{\epsilon_r+1}{2}$. In order to compare analytical results with experiment and the simulation we assume that the speed of light $c$ appearing in Eq. (\ref{Frequency}) is reduced by the square root of the effective dielectric constant. The analytically obtained resonance frequencies are in a good agreement with experiments results. For SR1 the relative deviation is 13\% for the first mode and 2-4 \% for the higher modes. For SR2 the relative deviation is 8\% for the first mode, 15\% for the second mode and 2-7 \% for the higher modes. The deviation is related to details of the experimental setup, different for SR1 and SR2, not taken into account in the model. Moreover, the difference in the resonance frequencies may be caused by the fact the resonators used in the experiments are not as densely packed, as it was assumed in the model.

\begin{figure}[tbp]
\includegraphics[width=3.2in]{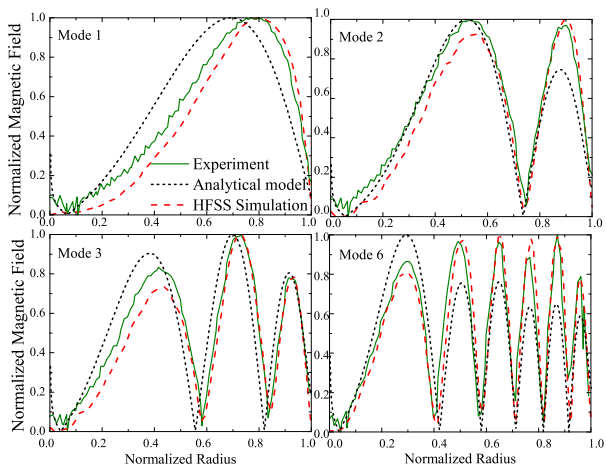}
\caption{Experimental (solid line), theoretical (dotted line), and numerically simulated (dashed line) amplitudes of the RF magnetic field versus radial coordinate of the spiral. The data are the normalized radial component of the magnetic field for the first, second, third, and sixth modes of copper spiral resonator (SR1). The radial component of RF magnetic field is measured with a network analyser via a mobile probing loop (as in Fig.\ref{ExpSimScheme}). The loop diameter is smaller than the spiral step, its plane is perpendicular to the radius, providing sensitivity to a radial component of the RF magnetic field. Also, the distance from the probing loop to the spiral surface is small (0.3 mm), compared to the spiral line width. Naturally, in the vicinity of the surface the radial component of the RF magnetic field profile is very similar to the profile of the RF current. The dotted line corresponds to the analytically obtained function $B_r(r, z)$ with $z=0.3$ mm of the perfect conductor spiral. The dashed line corresponds to the radial component of the magnetic field obtained by the HFSS simulation of the superconductive spiral on the substrate. Simulated magnetic field is taken at $z=0.3$ mm as well. The SR1 has 23 turns with the period $d=0.7$ mm and with external radius $R_e=16.25$ mm.}
\label{MagnField}
\end{figure}

Using Eqs. (\ref{sin_solution}) and (\ref{A-solution}), we obtain the RF current distributions in resonator SR2. These current distributions for four modes ($n=1, 2,3,$ and $6$) are presented in Fig. \ref{Currents} (dotted lines). One can see a good agreement in locations of the minima and maxima, and in the curve shape between experimental (solid lines) and HFSS simulated (dashed lines) RF current profiles.
\begin{center}
\begin{table*}
\centering
\begin{tabular}{|C{0.5 in}|C{1 in}|C{1 in}|C{1 in}||C{1 in}|C{1 in}|C{1 in}|}
\hline
    &\multicolumn{3}{c||}{Spiral Resonator 1 resonance frequency} & \multicolumn{3}{c|}{Spiral Resonator 2 resonance frequency}\\
\cline{2-7}
Mode number & Measured $f_{Exp}$, MHz & Analytical $f_{Anal}/f_{Exp}$ & HFSS simulated  $f_{HFSS}/f_{Exp}$ & Measured $f_{Exp}$, MHz & Analytical $f_{Anal}/f_{Exp}$ & HFSS simulated $f_{HFSS}/f_{Exp}$ \\
\hline
       1      &  80   &      0.87 &   1  & 128 & 1.08 &  1.14 \\
\hline
      2      &  180   &     0.96 &   0.99 & 299 & 1.15 & 1.03 \\
\hline
      3      &  268   &     0.92 &  0.99  & 465 & 1.05 &  1\\
\hline
      4      &  353   &    0.97  &   0.99 & 635 & 1.07 & 0.99 \\
\hline
      5      &  437   &     0.97 &  0.99  & 801 & 1.04 & 0.99  \\
\hline
      6      &  522   &     0.98 &   0.99 & 971 & 1.05 &  0.99\\
\hline
      7      &  607   &     0.97 &   0.99 & 1139 & 1.03 & 0.98\\
\hline
      8      &  693   &     0.98 &   0.99 & 1307 & 1.04 & 0.98\\
\hline
      9      &  779   &     0.97 &   0.99 & 1475 & 1.02 & 0.98\\
\hline
     10      &  866   &   0.98   &   0.98 & 1643 & 1.03 & 0.98 \\
\hline
\end{tabular}
\caption{Resonance frequencies of the SR1 and SR2 spiral resonators obtained experimentally, analytically and from the simulations.}
\label{freq_table}
\end{table*}
\end{center}\
In order to compare our analytical predictions with experiments carried out on resonator SR1, we calculated the magnetic field around the resonator. For this purpose, we approximate the spiral by the rings of variable radius $a$ with the current distribution $\psi(a)$ (Eq. (\ref{sin_solution})). Thus, the radial component of magnetic field is \cite{LL}
\begin{equation}\label{localBr}
\begin{gathered}
B_r(r, z)=\int_0^{R_e} da \frac{\psi(a)}{c}\frac{2z}{r\sqrt{(a+r)^2+z^2}}\cdot
\\
\Bigg[-K\left(\frac{\sqrt{4ar}}{\sqrt{(a+r)^2+z^2}}\right)+
\\
+\frac{a^2+r^2+z^2}{(a-r)^2+z^2}E\left(\frac{\sqrt{4ar}}{\sqrt{(a+r)^2+z^2}}\right)\Bigg],
\end{gathered}
\end{equation}
where $K\left(\frac{2r}{\sqrt{4r^2+z^2}}\right)$ and $E\left(\frac{2r}{\sqrt{4r^2+z^2}}\right)$ are complete elliptic integrals of the first and second kind. Using the corresponding $\psi(r)$ distribution, the radial component of magnetic field for the copper resonator (SR1) is calculated and is presented in Fig. \ref{MagnField} (dotted lines), also for all four modes. The magnetic field is calculated at the distance of $z=0.3$ mm from the plane of the resonator, corresponding to the setting of the probe loop used in experimental setup. The analytically obtained results are in a good agreement with the measured (solid lines) and with the HFSS predicted (dashed lines) mode profiles in terms of the locations of minima and maxima, and of the curves shape (Fig. \ref{MagnField}).

\section{Experiment}

Our electrodynamic analytical  model of the resonance modes of a spiral resonator has been verified through the direct measurement of resonance frequencies, the local RF currents and the RF magnetic fields of the spiral resonators. First spiral resonator (SR1) is fabricated as copper printed circuit board (PCB) on a hydrocarbon ceramic laminate substrate (RO4350B) with dielectric constant $\epsilon_r = 3.48$ and thickness 0.765 mm. The copper trace width is $0.3$ mm. The second resonator (SR2) is made from superconducting Nb thin film. SR2 is fabricated by lift-off photolithography on a Si substrate with dielectric constant $\epsilon_r = 11.45$ at 4 K temperature.\cite{Krupka} The width of the Nb trace is 10 $\mu$m and the gap between the adjacent turns is 10 $\mu$m wide.

The resonance frequencies of the both spiral resonators SR1 and SR2 are detected in RF transmission measurements in a way similar to Ref. \cite{Ustinov} and  the results are presented in the Table \ref{freq_table}. The two different experimental techniques  are used to study of electrodynamic properties of resonators.  The SR1 is studied by specifically designed MPS allowing one to measure the RF magnetic field distribution around the resonator. At the same time the superconducting resonator SR2 is studied with LSM, allowing one to obtain directly the RF current distribution across the spiral.

\begin{figure}[tbp]
\includegraphics[width=3.2in]{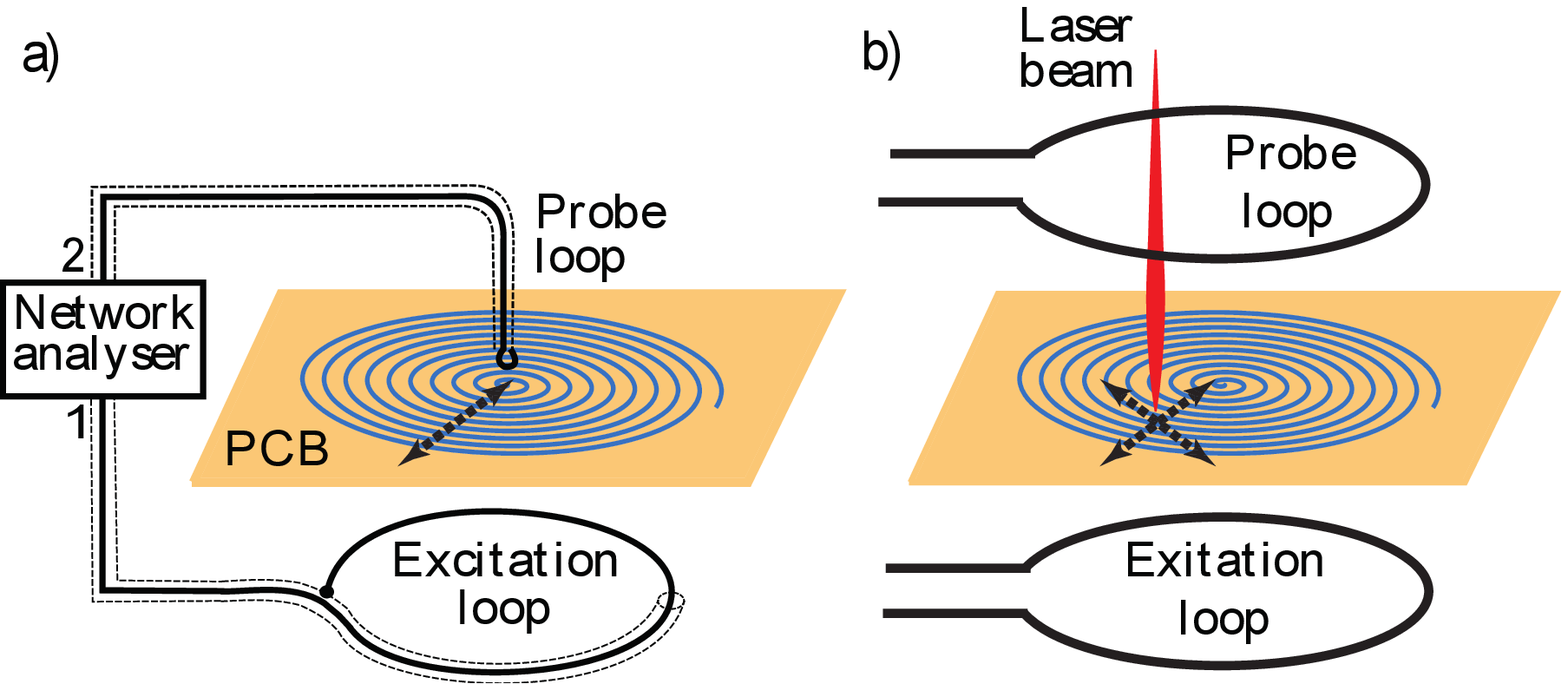}
\caption{Experimental setup for detecting the profile of the standing wave of the inner modes of a spiral resonator.\\
a) The sketch for RF magnetic probe (MPS) measurement. The probe loop is oriented perpendicular to the radius of the spiral to sense mostly the radial component of the RF magnetic field. The arrowed line represents trajectory (radius) along witch the probe loop can be moved by the motorized drive. \\
b) The sketch for laser scanning measurement. The laser beam is scanned over the superconducting spiral and perturbs the superconducting state in small area of the resonator. The enhanced concentration of normal electrons (quasiparticles) leads to a change of the transmission of the RF signal through resonator. The induced variation of the transmission coefficient $S_{21}$ corresponds to the squared amplitude of local RF current in the probed area of the circuit.\cite{LSM1}}
\label{ExpSimScheme}
\end{figure}

The radial component of the magnetic field near the surface of the SR1 is measured with MPS at room temperature using a small loop antenna as a probe. A sketch of the MPS experimental setup is shown in Fig. \ref{ExpSimScheme}a. Here, in order to excite the resonator, a 32 mm diameter shielded excitation loop is used as an RF field source. The loop is made of the 2 mm semi-rigid 50 Ohm coaxial cable and is positioned at 30 mm below the spiral resonator, far enough to ensure the weak coupling.

The radial component of the RF magnetic field near the surface of SR1 resonator is measured with loop antenna of about 0.5 mm in diameter formed at the end of 0.5 mm diameter semi-rigid 50 Ohm coaxial cable. This probe is placed at a distance of about 50 $\mu$m from the surface of the spiral and its plane is oriented perpendicular to the radius of the spiral. Thus, the full distance from the center of the probe loop to the surface of the spiral is about 0.3 mm. A motorized linear motion actuator is used to move the probe loop along the radius of the spiral and thus to measure the magnetic field spatial distribution. The movement step is 0.135 mm, small enough to resolve the RF magnetic field (current distribution) of the individual turns. Excitation signal is generated and probe signal is measured using a vector network analyzer. Both the motorized drive and the network analyser are controlled by PC. In our experimental setup, the $S_{21}$ transmission coefficient is proportional to the amplitude of RF voltage generated in the probe loop (while the excitation amplitude is constant), and thus proportional to the local amplitude of the magnetic flux.

The measured spatial distributions of the radial component of the RF magnetic field for four modes ($n=1,2,3,$ and $6$) are presented in Fig. \ref{MagnField} (solid lines). For each mode the curves are obtained by combining measurements along four orthogonal radial directions, allowing for complementing the data.


A low-temperature LSM\cite{LSM1} is used to examine in-plane $x-y$ distribution of RF microwave currents in superconducting resonator SR2. A number of specific schemes for the LSM electronics designed for the different detection modes have been published elsewhere \cite{LSM2, LSM3, LSM4, LSM5} and it is not a subject of discussion here. The LSM uses the principle of the point-by-point $x-y$ scanning of the superconducting planar structure by a sharply focused laser beam (See Fig. \ref{ExpSimScheme}(b)). Here, we use a setup that is specially designed for imaging of current distribution in superconducting RF samples with in-plane dimensions up to $10\times10$ mm. Our LSM spatial resolution is limited by the capabilities of objective lens focusing optical beam to about 5  $\mu$m. The focused beam acts as a local heat source in any point of the optical raster in the plane of the studied device. The power of the laser beam is set to about 10 nW (at 670 nm light wavelength). The induced perturbation is low enough not to change significantly the RF current distribution, while keeping LSM photo-response (PR) detectable. The intensity of the laser light is modulated in amplitude at the typical frequency of about $f_m\sim100$ kHz in order to increase the signal-to-noise ratio by using a lock-in detection technique. The modulation helps reducing the size of the thermal perturbation area proportional to $1/\sqrt{f_m}$.\cite{LSM1}

\begin{figure}[tbp]
\includegraphics[width=3.2in]{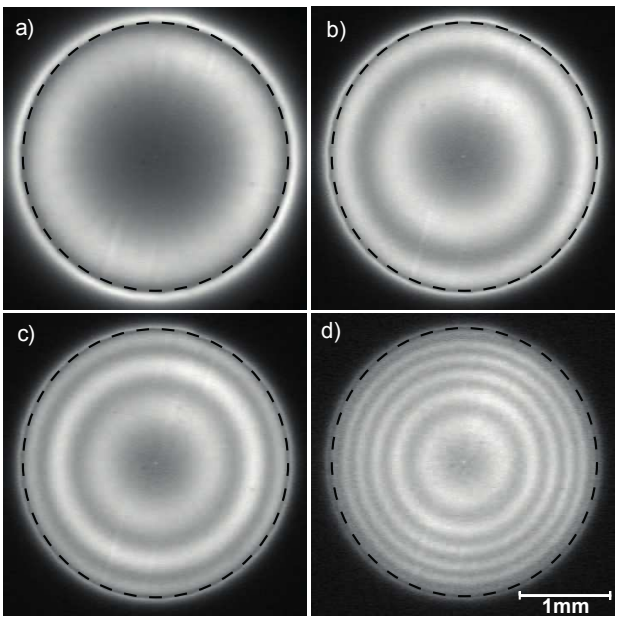}
\caption{The measured standing wave patterns of RF current for the first, second, third, and the sixth inner modes of Nb superconducting spiral resonator (SR2). The images are obtained with cryogenic Laser Scanning Microscope (LSM). Higher level of RF current is given in a lighter grey scale; the edge of Nb spiral is marked by a dashed line. The frequencies of the resonances in resonator SR2 are listed in Table \ref{freq_table}. Note a smearing effect of the light absorption in the spiral substrate, giving a halo outside of the spiral structure. The extracted radial profiles of the RF current of the modes are also smeared, see Fig. \ref{Currents}.}
\label{LSM}
\end{figure}

The laser scanning microscope images of the first, second, third, and sixth mode standing wave patterns of Nb superconducting resonator SR2 are presented in Fig. \ref{LSM}(a)-(d). The grey scale corresponds to the amplitude of photo response and can be interpreted as the squared amplitude of RF current flowing locally in the spiral (bright is large, dark is small RF current). The outer border of the spiral structure is given by the dashed line. One can see from the Fig. \ref{LSM}(a) that in the fundamental mode, there is the only one light-colored circle inside the resonator limits. The RF currents are lager in the middle part of the Nb resonator, and decay towards both ends of the spiral line. For the second mode (Fig. \ref{LSM}(b)), there are two rings of strong currents. Higher harmonics (Fig. \ref{LSM}(c),(d)) demonstrate a proportional increase in number of bright circles corresponding to local peaks in RF current. This confirms that the spiral acts as a distributed resonator with integer number of half-wavelengths of current at each eigenmode.
The cross-sectional radial profiles of standing waves are extracted from the images and are plotted by solid lines in  Fig. \ref{Currents}. One can note that all the profiles are smoothed off and, as a result, the measured current is not reaching zero at the RF current nodes. This can be explained by the smearing effect of the thermal spot, produced by laser beam when illuminating the Si substrate of the Nb resonator. The absorption coefficient in Si at 670 nm light wavelength is of about 65\%, and the calculated size of the thermal spot in Si substrate is larger than 20 $\mu$m (for T = 4 K and $f_m$=100 kHz). Thus, by extra heating through the substrate the laser beam generates additional LSM PR over 2-3 adjacent turns of Nb spiral. Note, that this experimental artifact does not affect the measured locations of the RF current nodes and antinodes.

\section{Simulation}

In order to verify the developed analytical model and to validate the experimental data, we use the ANSYS High Frequency Structural Simulator (HFSS).\cite{HFSS} The Driven Mode HFSS program is used for calculating the resonance frequencies and the shape of the standing waves of the resonant modes in both spirals used in our experiments. In HFSS calculations the planar spirals are assumed to be made of infinitely thin lossless metal layer situated at the substrate with appropriate dielectric permittivity $\epsilon_r$. The HFSS simulated circuit structure follows the experimental setup from Fig \ref{ExpSimScheme}b, where the spiral resonator sample is placed in between of the two weakly coupled magnetic loops. In order to ensure a weak coupling of the two loops, the inter-loop distance is set at four loop radii ($4R_e$). Also in HFSS model the two circuit terminals are inserted it a break in each of the coupling loops, and the resonance frequencies are determined from simulated terminal 1 to terminal 2 transmission ($S_{21}$) data.

The HFSS calculated resonance frequencies of SR1 are in a perfect agreement with experiment, remaining within 1\% of relative deviation for the first 10 resonant modes (Table \ref{freq_table}, column 4). Next, we simulate with HFSS the waveform of the standing wave of RF magnetic field for four resonance modes (n= 1, 2, 3, and 6) of spiral resonator SR1, as measured in experiment of Fig. \ref{ExpSimScheme}a. At the resonance frequencies we calculated the radial component of RF magnetic field along the radial line running from the spiral center, at a constant distance of 0.3 mm from the spiral surface (as in experiment Fig. \ref{ExpSimScheme}a). In this experiment the radial component of RF magnetic field is measured with a miniature loop probe at the same distance (solid line in Fig. 4). The HFSS calculated radial magnetic RF field component is plotted as well in Fig. 4 as a dashed line. The measured (solid line) and the HFSS predicted (dashed line) mode profiles are in a very good agreement in terms of the locations of minima and maxima along the radial axis, of the amplitude of the maxima, as well as in the curve shapes (Fig. 4). The analytical prediction has a similar precision, of about 3\% in resonance frequency, but requires a lower volume of calculations (Table \ref{freq_table}, column 3).

For the resonator SR2 made of superconducting Nb, the HFSS simulated resonance frequencies are listed in Table I, column 7. The precision of the HFSS prediction is very good, within 2\% of the measured values, except for the first resonant mode, where the deviation is 14\%. The deviation of the first mode resonance frequency may be related to the details of experimental setup, such as the metal parts of the cryogenic equipment, not taken into account in HFSS calculations. Next, we calculated with HFSS the waveform of the standing wave of RF current for four resonance modes (n= 1, 2, 3, and 6) of spiral resonator SR2. The value of the RF current in the spiral section is calculated applying Stoke's theorem (or Ampère's circuital law), as a line integral of the magnetic field around closed curve, encircling a section of conductive line of the spiral. This approach gave the best precision in calculation of the RF current, and is applied to 1800 points along the Spiral Resonator 2. The HFSS calculated profiles of RF current for the different modes in SR2 are given in Fig. 3 as dashed lines. In Fig. 3, one can note a very good mutual agreement in locations of the minima and maxima of experimental (solid line) and HFSS simulated (dashed line). The analytical prediction is giving a similar precision for the resonance frequencies, of about 4\%, except 15\% deviation at the second mode (Table I, column 6).

\section{Conclusion}

We developed an analytical model describing electrodynamics of a planar monofilar Archimedean spiral resonator of a finite length and verified it numerically and experimentally. We analytically obtain both the resonance frequencies and corresponding RF current distributions in spiral resonator. The set of resonance frequencies derived from our model is $f_n=f_1 n$, where $n=1,2...$, which is similar to resonances of a straight-line resonator with open ends. We note, that this simple relation for resonance frequencies is not trivial, and  depends on the details of the geometry of the spiral. For example, this relation fails to describe resonances in a ring-shaped spiral (with no central part).\cite{Anlage2,ourarticle}

Next, our analytic solutions are compared with experiment and direct numerical simulations. Experimental and simulated resonance frequencies follow the predicted relationship,  $f_n=f_1 n$, in agreement with analytical model (see the Table \ref{freq_table}). The difference between analytical and measured, or simulated results significantly decreases with increasing of the mode number. This can be due to the fact that, for lower modes, the electric and magnetic fields are wider spread in space than for upper modes. A widely spread RF field of lower modes interacts with surrounding parts of experimental setup, thus perturbing the spiral resonator eigenmodes and shifting their frequencies.
The shape of the standing waves predicted by analytical model is in a good accord with simulations and experiments (Fig. \ref{Currents}, \ref{MagnField}). The difference in positions of maxima and minima could be explained by the fact that the resonators used in experiment and in simulation are not as densely by packed as They are assumed in the model.

To summarize, we have studied theoretically (analytical calculations and numerical simulations) and experimentally the electrodynamic properties of the planar Archimedean spiral resonators of a finite length. By making use of a model of inhomogeneous alternating current flowing along the spiral resonator and of specific boundary conditions (see, Eq. (\ref{Condition})), we derived an integro-differential equation (\ref{Int-diff-Equation}) which, in turn, determines  the resonance frequencies and the corresponding RF current distributions in the spiral. Next, by applying a local approximation we solved Eq. (\ref{Int-diff-Equation}) and obtained the resonance frequencies and the corresponding current distributions of the Archimedean spiral resonator. A good quantitative agreement is found between the analysis, direct numerical simulations, and experiments performed with two spiral resonators made of different materials. Our results may find application in metamaterial design and in development of wireless power transfer (WPT) systems.

\section{Acknowledgments}

This work was supported in part by the Ministry of Education and Science of the Russian Federation Grant 11.G34.31.0062 and in the framework of Increase Competitiveness Program of NUST MISIS (K2-2014-015). N. Maleeva acknowledges the financial support from DAAD and Ministry of Education and Science of the Russian Federation through the Mikhail Lomonosov fellowship.

\appendix*
\section{}

In the Appendix, we present intermediate steps allowing one to obtain the integro-differential equation (\ref{Int-diff-Equation}). We use the following approximation: the wave vector $k$ is much smaller than a typical inverse size of inhomogeneities in current distribution $\psi(\rho)$, i.e. $k<<1/(R_e-R_i)$. Moreover, we consider a wide planar spiral,   i.e. $R_i \simeq 0$. With these approximations we have rewritten Eqs. (\ref{RproectionA-2}) and (\ref{ThetaproectionA-2}) as
\begin{equation}
A_r =\frac{\mu_0 I e^{i\omega t}}{(4\pi)^2} \int_{0}^{R_e} d \rho \psi(\rho) \int_{0}^{\infty}dx ~e^{-zx} J_1[\rho x]J_1[r x]
\end{equation}
and
\begin{equation}
A_\theta=\frac{\mu_0 I e^{i\omega t}}{(4\pi)^2 R_e \alpha} \int_{0}^{R_e} d \rho \psi(\rho)\rho \int_{0}^{\infty}dx ~e^{-zx} J_1[\rho x]J_1[r x].
\end{equation}

In order to obtain components of the vector potential in the plane of the spiral ($z=0$) lets use such an identity: $\left.A_r\right|_{z=0}=-\int_0^\infty dz \frac{\partial}{\partial z}A_r$.
\begin{equation}
\begin{gathered}
\left.A_r\right|_{z=0}=\frac{\mu_0 I e^{i\omega t}}{(4\pi)^2}\int_{0}^{\infty}dz \int_{0}^{R_e} d \rho \psi(\rho) \frac{1}{r^2}\cdot
\\
\cdot \int_{0}^{\infty}dx ~x e^{-\frac{zx}{r}} J_1[\frac{\rho}{r}x]J_1[x] \label{RproectionA-3}
\end{gathered}
\end{equation}
and
\begin{equation}
\begin{gathered}
\left.A_\theta\right|_{z=0}=\frac{\mu_0 I e^{i\omega t}}{(4\pi)^2 R_e \alpha}
\int_{0}^{\infty}dz \int_{0}^{R_e} d \rho \psi(\rho)\frac{\rho}{r^2} \cdot
\\
\cdot \int_{0}^{\infty}dx ~x e^{-\frac{zx}{r}}  J_1[\frac{\rho}{r}x]J_1[x]. \label{ThetaproectionA-3}
\end{gathered}
\end{equation}
Let us designate $\rho=R_e e^{-\tau}$ and $r=R_e e^{-\xi}$. Thus:
\begin{equation}
\begin{gathered}
A_r=\frac{\mu_0 I e^{i\omega t}}{(4\pi)^2R_e}\int_{0}^{\infty}dz \int_{0}^{\infty} d \tau \psi(\tau)e^{\tau} e^{-2(\tau-\xi)}\cdot
\\
\cdot \int_{0}^{\infty}dx ~x e^{-\frac{zx}{r}} J_1[e^{-(\tau-\xi)}x]J_1[x] \label{RproectionA-3}
\end{gathered}
\end{equation}
and
\begin{equation}
\begin{gathered}
A_\theta=\frac{\mu_0 I e^{i\omega t}}{(4\pi)^2 R_e \alpha}
\int_{0}^{\infty}dz \int_{0}^{\infty} d \tau \psi(\tau)e^{-2(\tau-\xi)} \cdot
\\
\cdot \int_{0}^{\infty}dx ~x e^{-\frac{zx}{r}}  J_1[e^{-(\tau-\xi)}x]J_1[x]. \label{ThetaproectionA-3}
\end{gathered}
\end{equation}

Next, we identify a kernel $K(\xi-\tau)$ as:
\begin{equation}\label{Kernel}
K(\xi-\tau)=e^{-(\tau-\xi)}\int_{0}^{\infty}dx ~x e^{-\frac{zx}{r}} J_1[e^{-(\tau-\xi)}x]J_1[x].
\end{equation}

Using Eq. (\ref{Kernel}) the components of the vector potential are expressed as:
\begin{equation}
A_r =\frac{\mu_0 I e^{i\omega t}}{(4\pi)^2R_e}\int_{0}^{\infty}dz \int_{0}^{\infty} d \tau \psi(\tau)e^{\xi} K(\xi-\tau)
\end{equation}
and
\begin{equation}
A_\theta =\frac{\mu_0 I e^{i\omega t}}{(4\pi)^2 R_e \alpha}
\int_{0}^{\infty}dz \int_{0}^{\infty} d \tau \psi(\tau) e^{\xi-\tau} K(\xi-\tau)
\end{equation}

Let us write the components of the electric field, i.e the angular component $E_\theta$:
\begin{equation}
E_\theta =-\frac{i\omega\mu_0 I e^{i\omega t}}{(4\pi)^2 R_e \alpha}
\int_{0}^{\infty}dz \int_{0}^{\infty} d \tau \psi(\tau) e^{\xi-\tau}  K(\xi-\tau)
\end{equation}
and
\begin{equation}
E_r=\frac{1}{i\omega\epsilon_0\mu_0}\left(-\frac{1}{R_e}e^\xi\frac{d}{d\xi}\left[-\frac{1}{R_e^2}e^{2\xi}\frac{d}{d\xi}(R_ee^{-\xi}A_r)\right]\right).
\end{equation}
Here, we have taken into account that $r=R_e e^{-\xi}$ and $\frac{d}{dr}=-\frac{1}{R_e}e^\xi\frac{d}{d\xi}$.

Using Eq. (\ref{Kernel}) the radial component of the electric field $E_r$ is written as:
\begin{equation}
\begin{gathered}
E_r=\frac{ I e^{i\omega t}}{i\omega\epsilon_0(4\pi)^2R_e^3}\int_{0}^{\infty}dz \int_{0}^{\infty} d \tau \psi(\tau)e^{3\xi}\cdot
\\
\cdot \left(K''_\xi(\xi-\tau)+2K'_\xi(\xi-\tau)\right)
\end{gathered}
\end{equation}

Finally, the Eq.(\ref{Condition}) is written as:
\begin{equation}
\begin{gathered}
\int_{0}^{\infty}dz \int_{0}^{\infty} d \tau \psi(\tau) \Big(e^{3\xi}K''_\xi(\xi-\tau)+
\\
+2e^{3\xi}K'_\xi(\xi-\tau) + \frac {\omega^2{R_e}^2}{c^2\alpha^2}e^{\xi-\tau}e^{-\xi}K(\xi-\tau) \Big)=0
\end{gathered}
\end{equation}
\\
\\
\\
\\

\end{document}